\documentclass[preprint,11pt]{aastex} 

\def\'#1{\ifx#1i{\accent"13\i}\else{\accent"13#1}\fi}

\def\alamenos#1{$^{-#1}$}

\def\be{\begin{equation}}

\def\diezala#1{10$^{#1}$}
\def\ee{\end{equation}}

\def\prommath#1{\langle #1\rangle}
\def\rhorms{\mbox{$\sigma_\rho$}}

\newcommand{\clumpfind}{{\tt clumpfind}}

\newcommand{\kms}{{\rm km~s}$^{-1}$}

\newcommand{\treceCO}{{$^{13}$CO}}

\begin{document}

\title{ Physical vs. Observational Properties of Clouds in Turbulent
Molecular Cloud Models} 

\author{Javier Ballesteros-Paredes and  Mordecai-Mark Mac Low}

\affil{Department of Astrophysics, American Museum of Natural
History\\ Central Park West at 79th Street, New York, NY, 10024-5192;
{\tt javierbp@amnh.org, mordecai@amnh.org}} 

\slugcomment{Draft date: \today}

\lefthead{Ballesteros-Paredes \&\ Mac Low}
\righthead{Cloud Properties}

\begin{abstract}
We examine the question of how well the physical properties of clumps
in turbulent molecular clouds can be determined by measurements of
observed clump structures. To do this, we compare simulated
observations of three-dimensional numerical models of isothermal,
magnetized,  
supersonic turbulence to the actual physical structure of the
models. We begin by determining how changing the parameters of the
turbulence changes the structure of the simulations. Stronger driving
produces greater density fluctuations, and longer wavelength driving
produces larger structures. Magnetic fields have a less pronounced
effect on structure, and one that is not monotonic with field
strength. Aligned structures are seen only with low-density tracers,
and when the intensity of the field is large. 
Comparing different regions with the same tracers (or
conversely, the same region with different tracers) can give
information about the physical conditions of the region. In
particular, different density tracers 
can help determine the size of the density fluctuations and thus the
strength of the driving. Nevertheless, velocity superposition of multiple
physical clumps can fully obscure the physical properties of those
clumps, and short wavelength (compared to the size of the region
under analysis) driving worsens this effect. 
We then compare Larson's relationships and mass spectra in physical
and observational space for the same structure dataset. We confirm
previous claims that the mean density-size relationship is an
observational artifact due to limited dynamical range in column
density: it is the inevitable consequence 
presence of a lower cutoff in column density. The velocity
dispersion-size relationship, on the other hand, is reproduced in both
physical and 
observed clumps, although with substantial scatter in the derived
slope, consistent with observations. Finally, we compute the mass
spectra for the models and compare them to mass spectra derived from simulated 
observations of the models. We show that,
when we look for clumps with high enough resolution, they both
converge to the same shape.  This shape appears to be log-normal,
however, rather than the power-law function usually used in the literature.

\end{abstract}


\section{Introduction}\label{intro}

This paper tries to understand whether
concentrations of emission in observations of position-velocity space
represent real clumps of gas in molecular clouds; whether the
properties inferred for the observed objects describe the
actual clouds; and what we can say about the mechanisms that drive
interstellar turbulence by observing molecular clouds with different
tracers.

Several papers have tried to address these points before.
\citet{AdlerRoberts92} show that, in a Galactic disk model, gas complexes
in a longitude-velocity diagram do not necessarily correspond to
real complexes. More recently, based on numerical simulations,
\citet{BVS99} showed that observed clumps frequently come from the
superposition of several physically disconnected regions in the line
of sight at the same radial velocity (see also \citet{Ostriker_etal01},)
but not necessarily at the same three-dimensional velocity.
Moreover, the morphology seen in observed
position-position-velocity maps seems to be more representative of
the spatial distribution of the velocity field in the line of sight
than of the distribution of the density field, as has been shown in 3D
numerical simulations by \citet{Pichardo_etal00}. Similarly,
\citet{Lazarian_etal01} shown that only the velocity field is
responsible for the structure existing in the channel maps of
observational data cubes.

If observed structures do not match actual structures, it is valid
to ask whether statistical properties of observed clouds, such as the
\citet{Larson81} relationships, are still valid for actual clouds. 
As \citet{Larson81} mentions, they may instead be artifacts of the
observations. In fact, \citet{Scalo90} argues that the mean
density-size relationship is an artifact of the limited dynamical
range of the observations (see also \citet{Kegel89}.) \citet{VBR97}
used two-dimensional simulations of the diffuse interstellar medium to
suggest that 
there is no relationship between mean density and size of 
physical clouds (defined as a connected set of 
pixels in their simulations), but that there is a
relationship, although with strong scatter, between the velocity
dispersion and size, a result confirmed by
\citet{Ostriker_etal01}. 

Although as we mentioned, several papers have studied the same
questions as the present paper, they addressed them only
partially, studying for example only objects in physical space
\citep{VBR97}, or in observational space
\citep{Ostriker_etal01}; or giving physical arguments to explain  
the observed \citet{Larson81} relationships (e.g., \citet{Larson81}
itself; \citet{Kegel89}; \citet{Scalo90}). In the present paper we use 
three-dimensional
numerical models of molecular clouds to directly compare structures in
simulated observations to the actual physical structures that
produce them.  Our goals are: to give observational predictions about
the nature of the sources of the turbulence; to disentangle what
observed properties are representative of the true physical entities;
and what are 
more likely to just be the result of superposition of multiple
structures along the line of sight.  Another way of asking this is,
what are the differences between the observational space and the
physical space? 

The plan of the paper is as follows: in \S\ref{simulationssec} we
describe the numerical models used here, and the methods used to
define clumps and to
generate simulated observations from model density and velocity cubes.
In \S\ref{role} we study the role of
the luminosity $L$, wavenumber $k$ and initial magnetic field $B$ in
the production of density fluctuations, and give observational
criteria to infer the properties of the sources of turbulence. We
discuss the superposition effects, and analyze the relative importance
of the parameters of the simulations in the superposition effects
(\S\ref{superposition}). Section~\ref{larsonsec} studies Larson's
relations (mean density vs.\ size, velocity dispersion vs.\ size and
mass spectrum) and discusses the importance of each one of them, by
comparing physical and observational space, and \S\ref{summary} draws
the main conclusions.

\section{Data Analysis}\label{simulationssec}

\subsection{Numerical Simulations}

In the present work we analyze driven, supersonic, hydro- and
magneto\-hydrodynamical (MHD) models described by
\citet{MacLow99}. Details of the models and forcing can be found
there. Here we just mention that the models are solutions of the
mass, momentum and induction equations using ZEUS-3D
\citep{StoneNorman92a,StoneNorman92b,Stone_etal96} at
a resolution of $128^3$ and $256^3$ zones. They include periodic
boundary conditions in each direction and an isothermal equation of
state, implying that the simulations are scale free. They are forced
at a characteristic scale given by the wavenumber $k$ and at a
constant energy injection rate $L$, with an initial magnetic field
intensity $B$.

Table \ref{simulations} shows the parameters of those simulations. The
first column shows the name of the run, following the notation in
\citet{MacLow99}\footnote{The first letter, M or H denotes MHD or 
hydrodynamical; the second letter gives the strength of the forcing,
ranging from {\em A} through two orders of magnitude in equal
logarithmic steps to {\em E}; the first number gives the value of the
driving wavenumber used; and in the magnetic case, the second number
shows the ratio $v_A/c_s$ of the Alfv\'en velocity to the sound speed,
with X representing 10.} In columns 2, 3 and 4 we show the driving
luminosity, wavenumber and initial intensity of the magnetic field
intensities. Column 5 shows the value of the root mean square of the
density fluctuations 
\begin{equation}\rhorms \equiv
\left[\frac{1}{N-1} \sum_{j=0}^{N-1} (\rho-\prommath
\rho)^2\right]^{1/2}. 
\end{equation} 

These simulations were performed to model the interior structure of
molecular clouds, at scales of 0.01-1~pc, densities of $n\sim
10^3$~cm\alamenos 3, and temperatures of $T = 10$~K.  Under these
conditions, the isothermal equation of state is a reasonable
approximation to the full thermodynamics of the cloud.  Other works
focussing on larger scales (10-1000~pc) have explicitly included
heating and cooling, a star formation scheme, and so forth
\citep{PVP95}, but 
these are not necessary to study the scales we are concerned with.
A bigger omission in our simulations is the lack of
self-gravity, which will play a role in the
production of dense cores
\citep{Klessen_etal00,Heitsch_etal01}. Nevertheless, we find the
non-self-gravitating simulations useful to study 
superposition effects, and their consequences in the statistical
properties observed for clumps. Even more, since our results are
consistent with other works that include self-gravity, we conclude that its
lack in our simulations is not relevant for the
conclusions stressed here.  A similar conclusion has recently been
reached by \citet{OKH01}, who studied wavelet transforms of
self-gravitating MHD turbulence and found that the overall structure
was decoupled from that of the densest cores.

\subsection{Simulated Observations}\label{observations}

We analyze the numerical data first by analyzing the physical
variables such as density, velocity, and magnetic fields themselves,
and second by simulating observations of the models. In previous work
\citep{BVS99,BHV99} we represented the line profiles of the
simulations as histograms of the velocity weighted by
density. However, those line profiles did not include the effect of
temperature in broadening the profiles, and did not account for
optical depth effects. Therefore we
calculate the radiative transfer here, assuming local thermodynamic
equilibrium (LTE) for the population of the molecular energy
levels, which is a sufficiently good approximation to study
qualitatively the effects of projection and superposition (velocity
crowding) of structure in the cloud.  In order to model
differences when observing the same region with different tracers, we
took into account that at low densities the molecules will
be underexcited. To do this we made the very simple assumption that the
lines are excited only 
if the density is above 650~cm\alamenos 3 for the \treceCO(1-0)
transition, 6200~cm\alamenos 3 for \treceCO(2-1), and
1.8$\times$\diezala 4 for CS(1-0) \citep{Rohlfs_Wilson96}. In
other words, we assume LTE if the local density is larger or equal to
these density thresholds for each transition.

Because the simulations are scale-free, the units of density,
temperature and length are arbitrary, but the included physics is
adequate to represent the behavior of molecular gas with length scales
between thousands of astronomical units for the densest cores, up to
several parsecs.  For the construction of the \treceCO(1-0), (2-1) and
CS(1-0) line profiles, we adopted solar abundances; and typical scales
for the density and length of $n=1000$~cm\alamenos3 
and $L=0.5$~pc, respectively. Since the simulations are isothermal,
we choose the temperature equal to 10~K. These choices produce regions
that are mostly optically thin in both \treceCO\ and CS.

In the production of line profiles, we fixed the number of
velocity channels at 32. This choice results in a
velocity resolution of 0.06~\kms\ for the less energetic simulations,
and of 0.2~\kms\ for the more energetic cases.  These resolutions are
comparable 
to the velocity resolution of typical observations of
molecular clouds at FCRAO which is 0.08~\kms, or in large-scale surveys
like
the full Galactic CO Survey by \citet{Dame_etal01} which reaches
resolutions as high as 0.26~\kms.

Since we know the detailed density and velocity distribution, and
assume a constant temperature, in order to determine
the emission from any given molecular transition we just have to
integrate the corresponding transfer equation,
\begin{equation}
\frac{dI_\nu}{ds} = \kappa_\nu \rho ( S_\nu-I_\nu)
\label{transfer}
\end{equation}
where $I_\nu$ is the intensity; $S_\nu$ is the source function, which
we assume to be a blackbody at $T=10$~K; $ds$ is the distance along the ray;
and $\kappa_\nu$ is the mass absorption coefficient of the
line, and has units of [cm$^2$ gr$^{-1}$]. In order to integrate equation (\ref{transfer}) along the
line of sight we solve, in practice, the following equations:
\begin{equation}
I_\nu = I
_{0} e^{-\tau_{\nu,{\rm tot}}} + B_\nu
\int_0^{\tau_{\nu,{\rm tot}}} e^{-\tau_\nu} d\tau_\nu 
\label{solution_transfer}
\end{equation}
and
\begin{equation}
\tau_\nu = \int_s^\infty \kappa_\nu\rho ds
\label{tau}
\end{equation}
where $I_{0}$ represents the background intensity and
$\tau_{\nu,{\rm tot}}$ is the total optical depth of the cloud, and
$\rho$ is the local mass density. 
Since the models have constant elements of volume (zones), the
integration over $s$ becomes a summation with constant step equal to the zone
size. The absorption coefficient $\kappa_\nu$ is 
\be 
\kappa_\nu = {A_{ij} c^2\over 8\pi \nu^2} {g_i\over g_j}
\biggl[1-\exp{\biggl(-{h\nu\over kT}\biggr)}\biggr] {n_j\over \rho} \Psi(v)
\label{kappa}
\ee
where $A_{ij}$ is the Einstein coefficient for the $i\to j$ transition
if the density is larger than the threshold density for excitation, 
and zero if the density is smaller; $g_i$ and $g_j$ are
the statistical weights of the upper and lower states respectively;
$n_j$ is the numerical density of molecules in the lower state; and
$\Psi(v)$ is 
the line profile, which is assumed to be thermal. These last two
parameters take the form 
\be
n_j = \rho {X_{\rm mol}\over m_0} g_j\exp{\biggl( -{E_j\over kT}
\biggr)} Q^{-1},
\ee
and
\be
\Psi(v)={c\over \nu}\biggl({m_{\rm mol}\over 2\pi
kT}\biggr)^{1/2}\exp\biggl( -{m_{\rm mol} (v-v_0)^2\over 2kT} \biggr),
\ee
where $m_0$ is the mean molecular mass of the gas (assumed to
be 2.5 times the mass of the hydrogen atom), $m_{\rm mol}$ is the mass
of the emitting molecule, $X_{\rm mol}$ is its molecular abundance
relative to the hydrogen (assumed to be solar), $Q$ is the partition
function, and $v_0$ is the velocity of the zone, which is known. The
molecular parameters are given in Table~\ref{molecules} 

\subsection{Physical and Observational Representation}

In order to represent and compare the physical and observational
spaces in two-dimensional plots we integrate over one of the dimensions. In
particular, when we integrate in the $x$ direction the
physical space, the corresponding observational cube is integrated in
the $z$ direction. This will allow us to compare, whether a clump in a
position-velocity diagram has a counterpart in the position-position
diagram. Then, when
we label a plot as $z-x(1-0)$ or $z-x(2-1)$, we show the
column density field integrated in the $y$ direction, after zeroing
all those points where the density is smaller than the density
threshold (see Table \ref{molecules}.)
The corresponding observational plot will be $z-V_x(1-0)$ or $z-V_x(2-1)$
respectively, and then represents
the simulated observation of the $z-x$ map as seen by an observer
located at the left of the plot, assuming \treceCO(1-0) or
\treceCO(2-1) emission as described above.

\subsection{Defining a Cloud}

The first problem that appears when we study the properties of clouds
and clumps is how to define them.  For instance, if we are interested
in studying large complexes and their substructure, we define a
complex as a connected set of points with intensity above an
arbitrary threshold \citep{Dame_etal86,VBR97}.
When we work in physical space, the ``intensity'' is the actual
density. If we work in the observational space, the
``intensity'' is the brightness temperature. If
instead, we are interested in studying individual clumps, we define a
clump as a connected set of points below a local maximum following the
intensity {\it only} downwards until the threshold is reached. This
scheme has been implemented by 
\citet{Williams_etal94} in a code called \clumpfind. A
single large cloud with internal structure in the first scheme may be
counted as several smaller clumps in the second scheme. Other schemes
are also used.  For example, one can assume that clumps have a
Gaussian distribution of intensity \citep{StutzkiGunsten90},
so that a clump defined as single by \clumpfind\ can be decomposed
into several Gaussian substructures.

In the present work we will use \clumpfind for simplicity.
Differences between methods will be discussed in future
contributions, but it is worth noting that, as \citet{Sheth_etal00}
has shown, these methods can not identify clumps larger than a few
resolution elements.

\section{Results}\label{results}

\subsection{Role of the parameters}\label{role}

We analyze how the parameters of the simulations (driving
strength and wavelength, magnetic field strength) affect the structure
in both observational and physical
space. In particular, we attempt to give observational
criteria to help determine the equivalent parameters of the physical
mechanism driving the observed turbulence. 

We begin by comparing models with the same driving
wavenumber and magnetic field, but different driving strengths. In
Table~\ref{simulations} we see that the larger the driving strength
$L$, the larger the density fluctuations \rhorms. This occurs because
the rms Mach number in the simulations depends on the driving strength
and the wavenumber as $M\propto (L/k)^{1/3}$ \citep{MacLow99}, and the
density contrast across an isothermal shock depends on the Mach number
as $\delta\rho/\rho \propto M^2$.  

Figure~\ref{maps_l} shows maps of both the physical column
density and the observed emission
for two hydrodynamical models with the same wavenumber
($k=8$), but driving strength different by a factor of 100: model HA8
($L=0.1$) and model HE8 ($L=10$). The main difference between these
extreme cases is  
that HA8 has larger column densities of low-density tracers,
and lower column densities of high-density tracers. Another clear
difference is that in high-density tracers HA8 shows well-separated,
roundish cores, while HE8 shows a more filamentary structure with few
compact cores. 

These differences can be understood as follows. When the turbulence is
weakly driven (HA8), density fluctuations are small.  There will be few
regions with density below
the density threshold for the low density tracer (\treceCO(1-0)), so
most of the cube will emit in that line, but conversely only a few places
will have densities higher than the threshold for the high-density tracer
(\treceCO(2-1)), so only a small column density is seen in that line.
Conversely, for strong driving (HE8), the
fluctuations around the mean are larger, and more regions will have
densities either below the threshold for \treceCO(1-0) or above the
threshold for \treceCO(2-1), so that the mean
column density for \treceCO(1-0) is lower, but the mean column density
for \treceCO(2-1) is larger.  Stronger driving leads not just to
higher densities, but more extreme fluctuations, as indicated by a
greater fraction of material at both high and low densities.
%

Since 
structures in the simulations are produced by the convergence of
turbulent flows, the smaller the driving wavenumber, the larger the
dominant structures.  We can see this effect in Figure~\ref{maps_k},
where we show four maps for HC2 and four maps for HC8, both with
the same driving strength ($L=1$), but different wavenumber ($k=2$
and~8 respectively). From these plots we can see that the density
structure is more homogeneously distributed in the case of high
wavenumbers (HC8), in both physical and observational spaces.
On larger scales, this suggests that {\it large molecular clouds might
be the result of turbulence driven at large scales in the diffuse
interstellar gas (see, e.g., \citet{BHV99, HBB01})
, and not the agglomeration of
small clouds by the action of the self-gravity}, as in some
models of molecular clouds (e.g.,
\citet{Scoville_Hersh79, Kwan79, Norman_Silk80})


The effect of magnetic fields in our models is consistent with
previous work using simulations of the diffuse interstellar medium by
\citet{PVP95}, who 
showed that weak magnetic fields decrease the size of the density
fluctuations compared to the hydrodynamic case. For intermediate
magnetic fields, the density fluctuations may increase above the
hydrodynamic and weak field cases, and finally, for very strong
magnetic fields, the density fluctuations decrease again. (The actual
diagnostic they used, as shown in their Fig.~8, is the star formation
rate as a function of the magnetic field strength; but in their scheme
the star formation rate depends directly on the size of the density
fluctuations). 

In Figure~\ref{magnetic} we demonstrate this behavior. The simulations
used have intermediate driving strength ($L=1$) and wavenumber
($k=4$), and three different initial magnetic field strengths of $B=0$
(HC4), $B=0.1$ (MC41), $B=0.5$ (MC45).  We can understand the
dependence of the density fluctuations on the magnetic field strength
as follows: weak magnetic fields are strongly tangled by the flow, so
that they have a nearly isotropic magnetic pressure that prevents
compressions and large density fluctuations, while stronger fields
have an anisotropic magnetic pressure that allows larger compressions
along the mean field lines, and thus larger density fluctuations.

Magnetic fields do not noticeably affect the structure of the maps
unless they are strong enough to introduce some anisotropy. For
example, in Figure~\ref{maps_b_10} we show \treceCO(1-0) maps of
three different simulations, integrated parallel ($z$), and
perpendicular ($x$) to the initial field direction. From bottom to
top, the relative importance of the magnetic field compared to the
luminosity of the simulation increases. We show: ME21 (large driving
strength, small wavenumber and 
small magnetic field); MA81 (low driving strength, large wavenumber and small
magnetic field); and MA4X (small driving strength, intermediate
wavenumber and large magnetic field).  Only the strong field case MA4X
(top panels) shows a morphology different than the hydrodynamic case.  In this
case, a marked anisotropy is visible, with the projection
perpendicular to the magnetic field (upper left panel),
showing low-density structures aligned mainly in the direction of the
field lines
while the projection along the field line (upper right panel)
presents regular structures, with no preferred
direction. Interestingly, in a higher density tracer, there is no
clear morphological difference between projections along or
perpendicular to the initial field (see Fig~\ref{maps_b_21}.)

The maximum values of the column density are similar in both physical and
observational projections for small magnetic field
intensity. Nevertheless, in 
the strong-field case (MA4X, see upper panels in Fig.~\ref{maps_b_10}
and~\ref{maps_b_21}) 
the column density integrated perpendicular to the mean field 
(right panels) is higher than the
column density integrated parallel to it (left panels), as can be seen
in the grayscale bars.
This effect can be explained in the same way as that shown in 
Figure~\ref{magnetic}: strong magnetic fields only allow compressions
parallel to the field. Therefore, when we observe along the magnetic
field lines we do not see high column density contrasts, while when
we observe perpendicular to the field lines, we see higher contrasts
in the compressions along
the field lines.
As a summary, {\it strong fields can produce anisotropy and compressed
structures perpendicular to the field lines, but weaker fields merely
reduce the maximum density contrast somewhat.} 

\subsection{Superposition Effects}\label{superposition}

Velocity crowding contributes substantially to the generation of
clumps in observational space, so observed clumps frequently
contain emission from physically separated regions
\citep{BVS99,Pichardo_etal00,Ostriker_etal01,Lazarian_etal01}. We 
demonstrate this effect using a typical MHD simulation: MC41. We
choose this run because it has intermediate values of driving strength
($L=1$) and wavenumber ($k=4$), and it clearly shows the effects of
superposition.

Figure~\ref{map_mc41} shows physical and observed maps for
\treceCO(1-0) and (2-1). We see that clumps in real space (letters A,
B, and C in panel $y-x(1-0)$) do not necessarily have a counterpart in
observational space.  Clumps in observational space, on the other
hand, (letter D in panel $V_y-x(1-0)$) do not necessarily come from isolated
regions in real space, but have contributions from many different
regions along the same line of sight. In Figure~\ref{map_mc41} we plot
dotted lines that show the places where the emission of a physical
clump lies in observational space, and where the emission from an
observational clump is generated in physical space. For reference, we
use the same lines in the \treceCO(1-0) map as in the \treceCO(2-1)
map. 

For the same data set we made another set of maps using higher density
tracers. Figure~\ref{mc41_CS10} shows the emission of CS(1-0). The same
lines are shown as in Figure~\ref{map_mc41}, to indicate the location
of the large \treceCO(1-0) core. In this case, the illusory
observational clump D has almost disappeared, and the physical clump A
is observed more clearly in its observational counterpart. We also
note that the position that we would attribute to the cores in
different tracers is slightly different.

In order
to quantify the effects of superposition, we now
compute the typical number $N_{\rm clump}$ of physically 
disconnected regions that contribute to the emission of a single
observed clump. To do so, we developed an algorithm that brackets each
observed clump in position and velocity and calculates the number of
connected regions in physical space within the bracketed region in
phase space.

In Figure~\ref{subclumps}{\em a} we show the influence of driving strength
and wavenumber on the amount of superposition.  We find that
high-density tracers behave differently than low-density tracers.  In
the case of a high-density tracer such as CS, $N_{\rm clump}
\sim$~2--3, and is almost constant for different driving strengths and
wavenumbers. On the other hand, in the case of a low-density tracer
such as \treceCO, which emits from most of the cube, superposition
increases with driving wavenumber. This is a natural consequence of
the way we forced the models, since short-wavelength driving produces
multiple physical regions at the same velocity.  There is also a
marginal tendency for strongly-driven models to exhibit more
superposition than weakly-driven ones. This occurs because the greater
velocity dispersion in the strongly-driven model produces clumps with
greater internal velocity dispersion that more easily overlap.  We conclude
that higher-density tracers will better disentangle velocity crowding,
as is shown by the results using CS(1-0), though still by no means
completely. 

In Figure~\ref{subclumps}{\em b} we show the relative influence of the
magnetic field by examining $N_{\rm clump}$ as a function of the ratio
of the driving strength to the initial magnetic field strength
$L/B$. We find that the amount of superposition is a bit larger when
we observe along the mean field direction ($z$) than when we observe
perpendicular to it. The effect is small, but present, and it is a
consequence of the greater compressions parallel to the field lines
producing more condensed clumps with smaller cross-sections for
superposition when observed perpendicular to the field lines. 

A final word concerning superposition is in order: although it is true
that high spatial resolution helps to segregate large clouds into their
clumps, high velocity resolution does not necessarily help to
disentangle superposition. It only helps if
there is a monotonic velocity gradient along the line of
sight, such that two clumps with similar velocities 
merge at low velocity resolution, and can be separated 
at higher velocity resolution.  If clumps are produced in
regions of converging flow in a larger turbulent region, however, as is likely
the case in real clouds, then 
the velocity gradients will not be
monotonic, and physically disconnected regions 
will often share the same line-of-sight velocity, so increasing
velocity resolution will not help.

As a summary, we conclude that velocity crowding is likely to always occur,
implying that cores in observational space are not single entities in
real space, and that it is larger for low-density tracers, large
driving wavenumbers, and when we look along the mean field lines. 

\section{Larson's Relations in Numerical Simulations}\label{larsonsec}

If observed clumps have
contributions from multiple physical regions, it is
important to understand whether the relationships reported for
observed clumps in molecular clouds have anything to do with the
properties of the actual physical clumps. 

\citet{Larson81} studied the dependence on size of the mean density,
velocity dispersion, and mass spectrum of the clouds in a sample of
observational data taken from the literature. He found 

\begin{equation}
 \prommath \rho \propto R^\alpha,
\label{rhomean}
\end{equation}

\begin{equation}
\delta v \propto R^m,
\label{deltav}
\end{equation}
and

\begin{equation}
{dN\over d \log{M}} \propto M^\gamma.
\label{espectro}
\end{equation}

The most commonly quoted values in the literature are $\alpha\sim -1$,
$m\sim 1/2$, and $\gamma\sim -0.5$. In particular, values of $\alpha=-1$ and
$m=1/2$ have been attributed to virial equilibrium\footnote{What is
often called virial 
equilibrium should actually be called energy equipartition, since (a)
observations are not able to measure all the terms in the virial
equation, such as the surface and time-derivative terms, to prove
virial equilibrium, and (b) it has been shown that clouds in
(two-dimensional) numerical simulations of the diffuse interstellar
medium are in energy equipartition \citep{BV95} but not necessarily in
virial equilibrium \citep{BV97}.}
\citep{Larson81,MyersGoodman88,VS_etal01}. However, there is some
discrepancy in the values reported \citep{Carr87,Loren89}, especially
for high latitude clouds, where the velocity dispersion-size
relationship has been questioned \citep{Larosa_etal99}. 

\citet{Larson81} himself already suggested
that the relationships he identified might be due to the limitations of
the observations, but it was \citet{Kegel89} who first demonstrated
that the observed mean density-size relationship could be due to
observational effects, and that the observed and physical properties
such as radius or volumetric density may be quite different. 
\citet{Scalo90} showed that CO and extinction saturate at roughly the
same column density, which forms the upper envelope of the mean
density-size relationship. He also discussed how the
surveys are biased due to the limitations of instrumental sensitivity and
saturation.   \citet{VBR97} reproduced the mass spectrum, as well as
the velocity dispersion-size relationship (although with large
scatter) for clumps in physical space, using two-dimensional models.
They reported 
the lack of a mean density-size relationship, confirming numerically
the analysis by Kegel (1989, see also the discussions in \citet{Larson81}
and \citet{Scalo90}) in the sense that there are clouds with small sizes
and low column density that will be undetected in observational
surveys.  More recently, \citet{Ostriker_etal01} reported a
slope close to $-1$ for the mean density-size relationship derived
from simulated observations of their models\footnote{Actually, they
report a slope of 2 in the mass-size relationship, but the
interpretation and result are the same.}, and a very flat correlation
between velocity dispersion and size ($m\sim 0.1$). As they mentioned,
both results are a consequence of their clump finding method (based on
what they call regions of contrast), which samples the entire line of
sight.

None of these studies attempting to explain the origin of the \citet{Larson81}
relationships have compared simulated observations to the actual
physical density and velocity distributions for the same numerical
model.  We now do that, using the three-dimensional simulations of
molecular clouds that we have described above. 

We must first define the size of a clump, both in observational and in physical
space. For clumps drawn from simulated observations we take the
circular radius, defined as $R_{\rm circ} = (A/\pi)^{1/2}$, where $A$
is the projected area of the clump on the plane of the sky, for
comparison with WGB94.  For physical clumps we take the geometric
radius, defined as $R=(R_x^2+R_y^2+R_z^2)^{1/2}$. Other definitions
are equally valid, but can lead to differences of a factor up to half
an order of magnitude when comparing different size definitions, and
differences of as much as a factor of 100 when calculating mean
densities by dividing the total mass by the size\footnote{Assuming
LTE. Non-LTE
effects will also increase the errors \citep{Padoan_etal00,Ossenkopf_01}.}.

Finally, since the results discussed in the present section do not
appear to depend on the parameters of the runs we analyze or (in the
magnetic case) on the projection with respect to the mean magnetic
field, we only present results for run HC8-256 (see
Table~\ref{simulations}), an available high-resolution run ($256^3$
zones). We search for clumps in physical space using \clumpfind,
considering only regions with a density larger than 650~cm\alamenos 3,
the critical density for \treceCO(1-0) emission. The results do seem
to depend on the clumpfinding scheme used and on its parameters. For
example, the number and size of the clumps found in \clumpfind\
depends on the level 
of refinement used in \clumpfind: at high refinement, large clumps can be
split into multiple smaller clumps, changing not only the slope of the
mass spectrum, but also the dynamical range of size in the scaling
relations. 
 
\subsection{Mean Density-Size relationship}\label{larson_rho}

In Fig.~\ref{rhomeanfig} (upper panel), we show the mean density-size
relationship for clumps in three-dimensional physical space. We
note three  
points about this Figure. First, the relation between mean
density and size given by equation~(\ref{rhomean}) is not seen at all,
confirming the results of 
VBR97 from two-dimensional models.  Second, there is a minimum density
below which there are no 
clumps identified.  This minimum is just given by the density
threshold we used in \clumpfind. Third, even though the simulations
exhibit a large dynamical range in density ($\rho_{\rm max}/\rho_{\rm
min} \sim 3.5 \times 10^4$), the dynamical range in the mean
density-size relationship is small, because in constructing such a
plot we choose the clumps around the local density maxima.

In the middle and lower panels of Fig.~\ref{rhomeanfig} we show the mean
density-size relationship for clumps in simulated observational maps
of the same run, integrated along the $x$ axis. The middle panel is the
one obtained by using CS(1-0), and the lower panel is the relationship
obtained by using \treceCO(1-0).  The observed clumps do exhibit
approximately the relationship given by \citet{Larson81}, despite the
lack of correlation exhibited by the physical clumps in this model.
We conclude from this demonstration that the observed density-size
relationship (eq.~[\ref{rhomean}]) is an observational artifact.

Two mechanisms have been suggested to explain density-size
relationship with slope $\sim -1$ in observations. First, it might be
due to the selection effect described by \citet{Kegel89}: only clouds
with intensity exceeding the noise threshold determined by the
instrumentation will be detected, effectively setting a column density
cutoff, rather than the physical density cutoff imposed in the
physical density-size relationship. A constant column density cutoff
produces a cutoff with slope $-1$ in the mean density-size plane
(middle and lower panels), just as the constant physical density
cutoff in physical space produces a flat cutoff.

Second, it might be due to the limited dynamical range of the
observations \citep{Scalo90}. To understand this, lets consider the
simulations: if we 
had clumps with constant column density $N$, we would infer a mean
density $\prommath{\rho(R)} = N R^{-1}$, giving exactly a slope of
$-1$.  In the simulation analyzed here we do not have a constant
column density, but the column density varies by only a factor of 30
(see Burkert \&\ Mac Low 2001), far less than the variation in
physical density.  Even worse, from this narrow distribution, the
selected clumps have an even narrower column density distribution: on the
one hand, as we already mentioned, there is a column density cutoff
given by the minimum intensity used in \clumpfind; on the other hand,
even the emission of the brightest clump does not come from the whole
line of sight, and its column  density is not the largest column
density available in the simulation. The end result is that selected
clumps naturally end up with column density constant to within an
order of magnitude, consistent with the observed scatter in the middle
and lower panels of Fig.~\ref{rhomeanfig}.

Similar arguments can be applied to the observational data. For
example, when observing emission lines from molecules, there is
a limited dynamical range in density available in which any particular
tracer can be excited: it will not be excited at too low density, and
it will become optically thick at too high a
density \cite{Scalo90}. At IRAS wavelengths, \citet{Wood_etal94} claim
constant column densities in IRAS 
observations of dust, where the dynamical range of the data should
have allowed detection of any significant variations. However,
\citet{VBR97} showed that for IRAS observations, only the external
layers of the clouds are heated by the diffuse UV radiation, so that
the emission at 60 and 100~$\mu$m actually comes from surface
layers that are indeed expected to have nearly constant column density. 

Finally, we point out that deviations from the mean density-size
relationship have also been found in observations
\citep{Carr87,Loren89,Falgarone_etal92}, suggesting that with 
sufficient care, the trap of apparently constant column density can be
overcome. 

\subsection{Velocity Dispersion-Size Relationship}\label{vel_disp}

In Figure~\ref{deltavfig} we show the velocity dispersion-size
relationship for clumps in physical and observational space. In the
case of physical space (upper panel) the typical
velocity dispersion is of order of 0.2~\kms, a value that coincides
with the value of the fluctuations of the overall velocity field for
the run analyzed here (HC8-256). The slope fit to the data is
0.3~$\pm$~0.38, smaller than the value of 1/2 suggested for
compressible turbulent behavior (see, e.g., VBR97), but with a strong
scatter. In fact, in order to reduce substantially the uncertainty in
the determination of the slope, we should have at least two orders of
magnitude of dynamical range in sizes if the scatter is about one
order of magnitude, as it is the case. But with \clumpfind, it is
difficult to have a large dynamical range (see, e.g., upper right
panel of Fig 7 in \citet{Williams_etal94}), since a reasonable
refinement of the intensity contouring makes large clumps split up
into several pieces \citep{Sheth_etal00}.

On the other hand, when we analyze observational data as in
Fig~\ref{deltavfig} (lower panel), additional problems play a role.
We now have the possibility that substantial superposition is
occurring, in particular for low-density tracers. If this is so, the
velocity dispersion of a single observed clump will reflect the
velocity dispersion of several, physically disconnected regions in the
same line of sight. This will also help to produce both large mean
values of $\delta v$ (0.8~\kms, compared to the 0.2~\kms\
corresponding to the mean velocity fluctuations in the simulation) and
low values of the slope in the velocity dispersion-size relationship
(Fig.~\ref{deltavfig}, lower panel).  

In contrast, for high density tracers (CS(1-0), Fig.\ref{deltavfig},
middle panel), we obtain a steeper relationship (0.46), and the mean
velocity dispersion drops back to $\sim 0.2$~\kms. Three reasons why
this occurs are: in the case of high density tracers there is weaker
superposition as shown in Fig.~\ref{subclumps}; clumps are
substantially smaller than in the low-density case, giving smaller
velocity dispersion; and the sizes span about one and a half orders of
magnitude, favoring disentanglement between the scatter and the actual
trend.

\subsection{Mass Spectrum}\label{masspecsec}

Now we turn to the clump mass spectrum.  Before showing the mass
spectra for the simulations, we have to think about the nature of the
turbulent density structure and how \clumpfind\ (or any other
clumpfinding algorithm) works. For instance, in a driven turbulent
medium, such as the simulations we examine here, we expect density
fluctuations at all scales above the dissipation scale.  It is not at
all clear when to stop counting substructure inside structure. In
other words, should we look for all the stones in the small hills in
every single mountain of the mountain chain, or only count the larger
mountains? If only the larger mountains, how do we define a lower
limit? 

Intuitively, we might think that the dependence of clump numbers on
\clumpfind\ parameters would be reduced in regions where clumps are
well defined and reasonably isolated, as in an isolated star-formation
region where cores have suffered some amount of collapse. To show that
this is {\it not} the case, we show in Figure~\ref{mapa_hc8} a contour
map of a slice through the center of model HC8. We use five and ten
isocontours between an arbitrary density threshold of $\rho_{\rm
max}/5$ and the maximum density $\rho_{\rm max}$. As we can see, the
density structures seem to be well-separated and reasonably well
defined. In Figure~\ref{closeup_hc8} we present an enlargement of the
lower-right corner of Figure~\ref{mapa_hc8}, again using five and ten
isocontours. We see that, even for the same density threshold, by
changing the number of isocontours we find five more peaks (see right
panel in Fig.~\ref{closeup_hc8}) that will result in five new clumps
in \clumpfind. This happens in just a small region containing what we
would think of as well-defined structures, at a single slice in
$z$. But in practice, we have 3D structures, so that every structure
is far more complicated than apparent at first glance. Even without
considering whether observational and physical clumps are the same or
not, the mass spectrum appears to be more an artifact of the manner in
which we count clumps than an actual physical characterization of the
structure, unless we count them exhaustively all the way down to the
numerical or physical dissipation scales below which there is no
longer well-defined structure. We note also that in
the case of self-gravitating simulations the result is the same
(Ossenkopf et al.\ 2001), and that similar results will hold for
observations: at low resolution the structures look soft, but once we
increase the resolution and the sensitivity, structure emerges inside
the structure, as is clear from Figure~4 in Dame et al.\ (2001). 

Preliminary work by \citet{Nordlund_02} shows that, by setting the
density threshold in physical space to the average density, a large
enough number of contour levels gives a log-normal mass spectrum,
independently of the number of contours used. We compute the mass
distribution using this prescription for run HC8 smoothed at a resolution
of $64^3$ for physical space, and $64^2\times 32$ velocity channels for
observational space in order to avoid excessive computation time and
memory usage by \clumpfind. The results are shown in
Fig.~\ref{masspec}. The upper panel corresponds to physical space,
and the lower panel corresponds to observational space, where instead
of density, we used intensity contouring. The results are clear:  for
large number of isocontours the shape of the mass spectrum
converges. Good agreement is observed in the large-mass end of the
distribution, although slower convergence is observed at the low-mass
end. Convergence to a factor of two for the full spectrum appears to
be reached for 64 or more contours.

In Fig.~\ref{lognormal} we compare the mass spectrum calculated in
physical space (bold line) to that from observational space (thin
line) using 512 isocontours.  The dotted
parabola shows a log-normal fit that reproduces fairly well the
large-mass end of the spectra. Such log-normal behavior of the
observed spectrum might explain observations obtained for dense cores
in the last few years. In particular, \citet{TestiSargent98} found a
power-law
mass spectrum with slope of $-2.1$ for star-forming cores in Serpens,
substantially larger than the standard $-1.5$ value. Moreover,
\citet{Motte_etal98} found a change in slope at low mass: the less
massive cores exhibit slope of $-1.5$, considerably shallower than the
$-2.5$ slope found for the more massive cores. Finally, we note
that in these works, the mass spectrum was obtained using dust
continuum observations with high signal-to-noise ratios. For present-day
molecular-line observations, typical signal-to-noise ratios of 10
or less do not allow small contour separation, since contouring
at less than the signal-to-noise ratio is meaningless
\citep{Williams_etal94}. Mass spectra derived from such observations
are likely to be far from physical (Fig.~\ref{masspec}).

\section{Summary}
\label{summary}

We analyze a set of three-dimensional MHD numerical simulations at intermediate
($128^3$ zones) and higher ($256^3$ zones) resolution described by
\citet{MacLow99}, in which 
the parameters driving strength ($L$), driving wavenumber ($k$), and initial
magnetic field ($B$) are varied. We show that:  

\begin{itemize}

\item{} Density fluctuations are primarily controlled by the
driving luminosity: the larger the luminosity, the larger the
fluctuations seen in the maps. Morphology of the
density structure, on the other hand, is primarily controlled by the
driving wavelength: 
the larger the driving wavelength, the larger the structures seen in
the maps, and vice-versa.  


\item{} The magnetic field only produces detectable effects in our
simulated molecular line 
observations if the fields are strong and in the plane of the sky, and
the structure is observed in low density tracers.  When observing magnetized
clouds, it is slightly easier to obtain larger column densities if the
field is in the plane of the sky than if it is perpendicular to it.

\item{} Simulated observational maps of the models show strong
superposition of structures along the line of sight. This implies that
single clumps in observational space are frequently the
superposition of multiple clumps in physical space.  The apparent
properties of observed clumps are poorly related to the physical
properties of their constituent objects if strong superposition
occurs.    

\item{} We explore the effect of the simulation parameters on the
generation of strong velocity crowding, finding that large driving
strength and large wavenumber tend to favor velocity crowding. So far,
the only more-or-less reliable method to ensure there is no
substantial superposition of clumps is to use high-density (close to
the maximum density) tracers. Nevertheless, even in those situations
superposition can occur.  

\item{} Larson's (1981) mean-density to size relationship does not exist in
physical space, but does occur in our simulated observations. It
appears to be an observational artifact, whose explanation we discuss.

\item{} Larson's velocity-dispersion to size relationship is
reproduced in both physical
clumps and simulated observations, though
with substantial scatter in the slope.  More dynamical range in the
simulations would probably help, although our clumps already span an
order of magnitude in size with half an order of magnitude in scatter.

\item{} The mass spectra converge to a particular shape only when we
use large number of contours (above 32 for the high mass end, but
above 64 for the whole spectrum), corresponding to an extraordinarily large
signal-to-noise ratio in the observations. Rather than a single power
law, they seem to follow a log-normal distribution.

\end{itemize}

\acknowledgements{We thank Paola D'Alessio, Jessica Moses and John
Scalo for careful reading of this manuscript, Paolo Padoan and \AA ke
Nordlund for 
pointing out the invariant, log-normal nature of the mass spectra, and
acknowledge 
useful discussions with Ralf Klessen, Volker Ossenkopf and John
Scalo.  Support for this work came from NASA Astrophysical Theory
Program grant NAG5-10103, CONACYT grant 88046-EUA, and NSF CAREER grant
AST-9985392. Computations were performed at the National Center for
Supercomputing Applications (NCSA), which is supported by the NSF, and
at the Rechenzentrum Garching of the Max-Planck-Gesellschaft. This
research has made use of NASA's Astrophysics Data System Abstract
Service.}

{}

\begin{figure}
\end{figure}

\begin{figure}
\caption{Physical ($z-x$) and observational  ($z-V_x$) maps for two models
with the same wavenumber ($k=8$) and no magnetic field ($B=0$), but
different values of the driving strength: (a) HA8 ($L=0.1$), and (b)
HE8 ($L=10$).  For weaker driving, the density fluctuations are small
and the entire cube contributes to the emission in low density tracers
(HA8 $z-x$(1-0) and $z-V_x$(1-0)), but only a few places contribute to the
emission in high density tracers (HA8 $z-x$(2-1) and $z-V_x$(2-1)). In
contrast, for stronger driving, pixels with density below the
threshold do not contribute, giving comparatively lower column density
values in low density tracers (HE8 $z-x$(1-0) and $z-V_x$(1-0)), but
substantially more regions are at high densities, giving higher column
densities in high density tracers (HE8 $z-x$(2-1) and $z-V_x$(2-1)).
\label{maps_l}}
\end{figure}

\begin{figure}
\end{figure}

\begin{figure}
\caption{Physical and observational maps for two runs with
the same driving strength ($L=1$) and no magnetic fields ($B=0$), but
large and small values of the wavenumber: (a) HC2 ($k=2$), and (b) HC8
($k=8$). The larger the wavenumber, the smaller the size of the
dominant scale for density structures in the simulation. 
\label{maps_k}}
\end{figure}

\begin{figure}
\caption{Root mean square of the density fluctuations (\rhorms) as a
function of the initial magnetic field strength of the
simulations. Low magnetic field strength prevents density fluctuations
more efficiently than either zero or larger-strength fields. 
\label{magnetic}}
\end{figure}

\begin{figure}
\caption{Physical ($z-x$(1-0) and $y-x$(1-0)) maps for large-scale,
large intensity 
driving and  
small magnetic fields (ME21, lower panels); small-scale, small
intensity driving and small magnetic fields (MA81, middle panels);
and intermediate scale, small intensity driving and strong magnetic
fields (MA4X, upper panels) integrated along the $y$ (left panels) or
$z$ (right panels), where the initial magnetic field runs parallel to
the $z$-axis. The relative importance of the magnetic field (compared
to the luminosity) increases
upwards. Column densities are similar in both projections
(perpendicular and parallel to the mean magnetic field lines), except
if the magnetic field is strong (upper panels). In this case,
projections along the magnetic field gives smaller column density than
the projection perpendicular to it. As explained in
the text, strong magnetic fields allow only compressions parallel to
the field, and then only when observing perpendicular to the field
lines we get high density contrasts.
\label{maps_b_10}}
\end{figure}

\begin{figure}
\caption{Same as Figure~\ref{maps_b_10} but for (2-1). Note that
the total column density is still typically larger when we integrate
perpendicular to the field lines than along them. Nevertheless,
differences in the morphology are more subtle, showing that the
alignment of the structures to the magnetic field occurs
preferentially at low densities. 
\label{maps_b_21}}
\end{figure}

\begin{figure}
\figcaption{ Physical and velocity space maps for \treceCO(1-0) and
(2-1) in run MC41 ($L=1$, $k=4$ and $B=0.1$). Clumps in physical space
(A, B, C) do not necessarily correspond to clumps in the observed velocity
space. Observed clumps in velocity space (D) are not necessarily
formed by emission from a single region in physical space. 
\label{map_mc41}}
\end{figure}

\begin{figure}
\figcaption{Physical and velocity space maps for the CS emission of run
MC41. Since CS traces higher densities, the clumps are more separated
than in the \treceCO\ emission shown in Figure~\ref{map_mc41}. Note
that the position of the cores A, B, C, D has changed
slightly respect to their position in Fig.~\ref{map_mc41}
\label{mc41_CS10}}
\end{figure}

\begin{figure}
	{f8b.eps}
\figcaption{(a) Mean number of physical regions contributing to each
clump in \treceCO(1-0) and in CS(1-0) ($N_{\rm clump}$)  versus
$\log_2{k}$. The different types of points denote different driving
luminosities $L$. Superposition accounts better for large wavenumbers,
low-density tracers, strongly driven simulations. (b) $N_{\rm clump}$
for models both parallel ($z$) and perpendicular ($x$) to the mean
magnetic field direction {\it vs} ratio of driving strength to initial
field strength $L/B$ for models MA4X (weak driving, strong field),
MC4X (stronger driving, strong field), MC41 (stronger driving, weak
field). Error bars show the standard deviation.
\label{subclumps}}
\end{figure}

\begin{figure}
\caption{Mean density-size relationship for physical clumps in physical
(upper panel); and simulated observational clumps in observational coordinates
(middle and lower panels). The dotted line has a slope of $\alpha =
-1$. In physical space we find no correlation, verifying the results
by VBR97, but nevertheless the simulated observations show such a
correlation, as found by Larson (1981) and many others. The selection
of two different density tracers was chosen to show that the apparent
correlation does not depend on the selection of the density
threshold.
\label{rhomeanfig}} 
\end{figure}

\begin{figure}
\caption{Velocity dispersion-size relationships for run HC8-256 in
physical space (upper panel), and observational space (middle and
lower panels) using CS(1-0) and \treceCO(1-0) respectively.  The
dotted line has a slope of 1/2, the expected value for a turbulent
medium dominated by shocks (see, e.g., V\'azquez-Semadeni et
al. 2000).  The solid line is the least squares fit to the data
points, with a slope $m$ and its uncertainty shown in each frame.  
\label{deltavfig}}
\end{figure}

\begin{figure}
\caption{Isocontours for the density field $\rho > \rho_{\rm max}/5$
at $z=64$ pixels using (a) five isocontours, and (b) ten
isocontours. Note that the clumps seem to be well-defined and
isolated. 
\label{mapa_hc8}}
\end{figure}

\begin{figure}
\caption{Enlargement of the lower-right corner of
Fig.~\ref{mapa_hc8}. We use again (a) five, and (b) ten
isocontours. Note that many new clumps appear with the increased
number of contours, even for well defined, isolated clumps.
\label{closeup_hc8}}
\end{figure}

\begin{figure}
\caption{Mass distribution of physical (upper panel) and observational
(lower panel) clumps in run HC8, computed using \clumpfind\ with
density threshold equal to the mean density (physical space) or mean intensity
(observational space), and varying the number of contours: 8 (dotted
line), 32 (thin, 
solid line), 128 (medium bold solid line), and 512 (bold solid
line). After $\sim 64$ contours (not shown here for clarity purposes),
the shape of the whole mass spectrum converges reasonably well. Below
that resolution, the final number of clumps, and especially their
distribution for the low-mass end varies strongly with the contouring
chosen.
\label{masspec}}
\end{figure}

\begin{figure}
\caption{Mass distribution (solid lines) for physical (thin) and
observational (bold) spaces using \clumpfind\ with 512 contours and a
threshold equals to the mean density or intensity, respectively. Note
that the mass spectrum inferred for observations and simulations
converges to the same shape, a log-normal distribution, as shown by
the dashed-line parabola.
\label{lognormal}}
\end{figure}

\begin{deluxetable}{rrrrr}
\tablecolumns{5}
\tablecaption{Properties of the simulations. \label{simulations}} 
\tablewidth{0pt}
\tablehead{
	\colhead{Name}
	&\colhead{$L$}
	&\colhead{$k$}
	&\colhead{$B$}
	&\colhead{\rhorms}
}
\startdata
	  {MA81}
	& {0.1}
	& {8}
	& {0.1}
	& {0.68}
\\
	  {MC81}
	& {1}
	& {8}
	& {0.1}
	& {1.21}

\\
	  {MC85}
	& {1}
	& {8}
	& {0.5}
	& {1.56}
\\ 
	  {MC45}
	& {1}
	& {4}
	& {0.5}
	& {1.83}
\\ 
	  {MA4X}
	& {0.1}
	& {4}
	& {1}
	& {1.27}
\\ 
	  {MC4X}
	& {1}
	& {4}
	& {1}
	& {2.21}
\\ 
	  {MC41}
	& {1}
	& {4}
	& {0.1}
	& {1.35}
\\ 
	  {ME21}
	& {10}
	& {2}
	& {0.1}
	& {2.32}
\\ 
	  {HC4}
	& {1}
	& {4}
	& {0}
	& {1.97}
\\ 
	  {HC8}
	& {1}
	& {8}
	& {0}
	& {1.40}
\\ 
	  {HC8-256}
	& {1}
	& {8}
	& {0}
	& {2.00}
\\ 
	  {HA8}
	& {0.1}
	& {8}
	& {0}
	& {.71}
\\ 
	  {HB8}
	& {0.3}
	& {8}
	& {0}
	& {1.05}
\\ 
	  {HD8}
	& {3}
	& {8}
	& {0}
	& {1.82}
\\ 
	  {HE8}
	& {10}
	& {8}
	& {0}
	& {1.97}
\\ 
	  {HE4}
	& {10}
	& {4}
	& {0}
	& {2.39}
\\ 
	  {HC2}
	& {1}
	& {2}
	& {0}
	& {2.21}
\\ 
	  {HE2}
	& {10}
	& {2}
	& {0}
	& {2.50}
\enddata
\tablecomments{All runs are at a resolution of $128^3$ except run
HC8-256, which is the equivalent to run HC8 but at a resolution of
$256^3$ pixels} 
\end{deluxetable}

\begin{deluxetable}{r c c c r r c}
\tabletypesize{\scriptsize} 
\tablecaption{ \label{molecules}}
\tablewidth{0pt}
\tablehead{	\colhead{Molecule}
	        &\colhead{Frequency}
		&\colhead{$A_{ij}$}
		&\colhead{Dipole Moment}
		&\colhead{$X_{\rm mol}$}
		&\colhead{Threshold density}
		&\colhead{References}
\\
	 \rm{}
	&\rm{(s\alamenos 1)}
	&\rm{(s\alamenos 1)}
	&\rm{(Debye)}
	&\rm{(relative to H$_2$)}
	&\rm{for excitation}
	&\rm{}
}
\startdata
	 \rm{$^{13}$CO(1-0)}
 	&\rm{$1.1\times 10^{11}$}
	&\rm{$7.6\times10^{-8}$}
	&\rm{0.112}
	& \rm{1.43 $\times 10^{-6}$}
	&\rm{650 cm\alamenos 3}
	&\rm{1,2}
\\ 
	 \rm{$^{13}$CO(2-1)}
	&\rm{$2.2\times10^{11}$}
	&\rm{$6.5\times10^{-7}$}
	&\rm{0.112}
	&\rm{1.43$\times10^{-6}$}
	&\rm{6200 cm\alamenos 3}
	&\rm{1,2}
\\ 
	 \rm{CS(1-0)}
	&\rm{$4.9\times10^{10}$}
	&\rm{$2.1\times10^{-6}$}
	&\rm{1.957}
	&\rm{1.0 $\times 10^{-10}$}
	&\rm{18000 cm\alamenos 3}
	&\rm{1,2}
\\ \hline

\enddata
\tablenotetext{1}{\citet{Rohlfs_Wilson96}}
\tablenotetext{2}{\citet{Gomez_DAlessio00}}
\end{deluxetable}

\end{document}